\documentclass[twocolumn,aps,pra]{revtex4-1}
\usepackage{color}
\definecolor{MyDarkBlue}{rgb}{0,0.08,0.60}
\usepackage{amsmath}
\usepackage{amssymb}
\usepackage{graphicx}
\usepackage{esint}
\usepackage{epstopdf}
\usepackage{amsfonts}
\usepackage{bm}

\newcommand{\beq}{\begin{equation}}
\newcommand{\eeq}{  \end{equation}}
\newcommand{\bea}{\begin{eqnarray}}
\newcommand{\eea}{\end{eqnarray}}
\newcommand{\beqn}{\begin{eqnarray}}
\newcommand{\eeqn}{\end{eqnarray}}
\newcommand{\bit}{\begin{itemize}}
\newcommand{\eit}{  \end{itemize}}

\newcommand{\gsim}{\mathrel{\hbox{\rlap{\lower.55ex 
\hbox{$\sim$}} \kern-.3em \raise.5ex \hbox{$>$}}}}

\begin{document}

\title{Optimal Uncertainty Relations for Extremely Coarse-grained Measurements}

\author{{\L{}}ukasz ~Rudnicki}

\email{rudnicki@cft.edu.pl}

\affiliation{Center for Theoretical Physics, Polish Academy of Sciences, Aleja
Lotnik{\'o}w 32/46, PL-02-668 Warsaw, Poland}

\author{Stephen P. ~Walborn}

\affiliation{Instituto de F{\'{i}}sica, Universidade Federal do Rio de Janeiro,
Caixa Postal 68528, Rio de Janeiro, RJ 21941-972, Brazil}

\author{Fabricio~Toscano}

\affiliation{Instituto de F{\'{i}}sica, Universidade Federal do Rio de Janeiro,
Caixa Postal 68528, Rio de Janeiro, RJ 21941-972, Brazil}
\begin{abstract}
We derive 
{two} quantum uncertainty relations for position and momentum coarse-grained measurements.  Building on previous results, we first improve the lower bound for uncertainty relations using the R\'enyi entropy, particularly in the case of coarse-grained measurements.  We then 
{sharpen} a Heisenberg-like uncertainty relation {derived previously in [Europhys. Lett. \textbf{97}, 38003, (2012)]}
that uses variances and
reduces to the usual one in the case of infinite precision measurements.
{Our sharpened uncertainty relation} is meaningful for any amount of coarse graining.  That is, there is always a non-trivial uncertainty relation for coarse-grained measurement of the non-commuting observables{, even in the  limit of extremely large coarse graining}.
     
\end{abstract}


\maketitle

\section{Introduction}

Uncertainty relations play a central role in quantum physics. Generally,
these are inequalities that limit the amount of information that can be obtained
about non-commuting observables for identically prepared systems. 
From a fundamental point of view,
the fact that measurements on quantum systems must obey uncertainty
relations distinguishes them from their classical counterparts. Since
all physical quantum states must obey them, uncertainty relations
provide a method to check the validity of an inferred quantum state
that was reconstructed from tomographic measurements. Uncertainty
relations also play an important role in applications in quantum information
science. In particular, the security of several quantum cryptography
protocols are founded on uncertainty relations \cite{reid00, grosshans04},
as are several tests for detection of quantum entanglement \cite{simon00,duan00,mancini02,nha08,walborn09,saboia11} and EPR-steering correlations \cite{reid89,wiseman07,reid10,walborn11a}.

{Historically, the most renowned uncertainty relation is: }
\begin{equation}
\sigma_{x}^{2}\sigma_{p}^{2}\geq
\frac{1}{4}|\langle[\hat x,\hat p]\rangle|^2=
\frac{\hbar^{2}}{4},
\label{Heisenberg-uncertainty-rel}
\end{equation}
{ for position $\hat x$ and momentum $\hat p$ observables. 
Its existence was first suggested in \cite{heisenberg},
and since then has been called the Heisenberg Uncertainty Relation (HUR). It was proved by 
Kennard in \cite{kennard}, and later Robertson  \cite{robertson} extended its validity 
to arbitrary pairs of observables $\hat A$  and $\hat B$.}
The HUR  \eqref{Heisenberg-uncertainty-rel}
applies to the variances, $\sigma_{x}^2 \equiv
\langle\hat x^2\rangle-\langle\hat x\rangle^2$ and 
 $\sigma_{p}^2 \equiv
\langle\hat p^2\rangle-\langle\hat p\rangle^2$, 
which quantify the uncertainty in position  and
momentum  measurements, respectively. Here $\langle \ldots \rangle \equiv \mathrm{tr}[\hat \varrho \ldots]$ is the expectation value when the system is described by the quantum state $\hat{\varrho}$.
This inequality follows from the fact that the complementary
operators do not commute: $\left[\hat{x},\hat{p}\right]=i\hbar$.
A number of additional uncertainty relations  for position and momentum have
been presented \cite{bialynicki75,deutsch83,partovi83,bialynicki84,kraus87,maassen88,bialynicki06,
vicente08,zozor11}.
 
{ It is important to distinguish the uncertainty relation in 
Eq.(\ref{Heisenberg-uncertainty-rel}) from the similar ones that appear
(\textit{i}) in the context of joint measurement of position and momentum
\cite{Arthurs-Kelly,Raymer1994},  and (\textit{ii}) in the error-disturbance  
process of position measurement \cite{ozawa,erhart}. 
In the case (\textit{i}) we have the {\it Heisenberg uncertainty relation for joint measurement}, and in  case (\textit{ii}) the {\it Heisenberg noise-disturbance uncertainty
relation}, which establishes a lower bound between the product of the estimate of a position measurement error and the estimate of the resulting disturbance in the momentum.  Ref.  \cite{ozawa} provides an analysis of the connection between these two types of uncertainty relations and Ref. \cite{watanabe} describes an alternative approach to the trade-off  relation between the measurement errors of noncommuting observables.
For case (\textit{i}), the joint measurement of position and momentum,} it was shown in \cite{Arthurs-Kelly} that, independent of the strategy used to perform the joint measurement, the  lower bound of the product of variances is incremented by a factor of two. This is due to the fact that joint measurement of the original non-commuting variables requires that each one must interact with different ``meter'' variables that have to commute themselves in order to be jointly measurable (in principle with arbitrary high precision). It is this coupling of the original variables with the meter variables that introduce the additional noise. 

\par
Uncertainty relations are generally interpreted to express limits
to the amount of information that one can obtain about complementary
properties of a quantum system prepared in a given quantum state, 
and thus naturally invoke the notion
of measurement. 
However, most uncertainty relations, such as \eqref{Heisenberg-uncertainty-rel},  assume perfect knowledge of the quantities involved,  which can only be obtained experimentally with infinite precision measurements.
However, in any experiment, measurements  are performed
with a finite precision, and thus any consistent uncertainty
relation applicable to ``real-world" scenarios involving statistics of measurement results 
should include
aspects of the measurement process.  Several authors have considered
finite precision, or ``coarse-grained'' measurements in the context of uncertainty relations involving
the Shannon \cite{partovi83,bialynicki84} and R{\'e}nyi entropy functions
\cite{bialynicki06}.  These relations are meaningful up to a certain amount of coarse graining, after which they are trivially satisified \cite{Rudnicki2011}.  
\par
The analysis of imprecision (coarse-grained measurement) in the context of  joint measurement of position and momentum (case (\textit{i}) above) was given in \cite{Raymer1994}, and   
a HUR-type uncertainty relation for joint measurement was obtained for sufficiently small values of the  resolution of the detectors. 
It is important to note that in this HUR inequality the joint measurement necessarily relates the widths of the resolutions of position and momentum measurements. 
The coarse-grained version of the HUR uncertainty relation in Eq.(\ref{Heisenberg-uncertainty-rel}) that arises not from joint measurement, but rather from the statistical results of measurement of position and momentum in an identically prepared system was recently obtained in \cite{Rudnicki2011}. This inequality, in contrast to the one in \cite{Raymer1994}, is indeed valid for arbitrary and independent values of the widths of the position and momentum detectors (although for large enough values of coarse graining the inequality is trivially satisfied).

It is generally believed, and taught in many quantum mechanics text
books \cite{peres95,ballentine98}, that the quantum nature of a system is not
observable when coarse-grained measurements are performed. This has been shown in the context of Bell's inequality violations \cite{peres92} and for the precession of a single spin-$j$ particle \cite{kofler07}.  Thus, it seems  natural to expect that uncertainty relations should at some point fail to be
significant with an increasing amount of coarse graining.   
{However, here} we derive new quantum mechanical uncertainty relations, and show through optimization that
coarse-grained measurements are \emph{always} limited by some quantum uncertainty
relation.  We consider the HUR and a family of uncertainty relations using
the R{\'e}nyi entropy under coarse-grained sampling. First, we show that one can construct reliably estimated probability distributions for coarse-grained measurements in order to obtain reliable uncertainty relations.
We provide an improved lower bound for the Bialynicki-Birula uncertainty
relations \cite{bialynicki06} using discrete R{\'e}nyi entropies, and use this
lower bound to derive a new Heisenberg-like uncertainty relation for
coarse-grained measurements.  By optimization, we provide a Heisenberg-like uncertainty
relation that is meaningful for \emph{any amount of coarse graining}.  That is, the amount of information
that can be obtained by coarse-grained measurements of non-commuting observables is always limited. We also prove that this optimized uncertainty relation reduces to the usual one in the limit of infinite precision measurements.
\par    
This paper is organized as follows. In sections \ref{sec:introduction}
and \ref{sec:probability} we introduce several definitions and previously
established uncertainty relations. Section \ref{sec:four} provides a method for constructing
continuous probability density functions from coarse-grained measurements.  
In section \ref{sec:entropic}
we derive an improved lower bound for the Bialynicki-Birula uncertainty
relation for discrete R{\'e}nyi entropies. We show that this relation is optimal
in the case $\alpha=1/2$, and provides an improvement for the case
of the Shannon entropy ($\alpha=1$) for large coarse graining. In
section \ref{sec:Heisenberg} we derive the HUR for coarse graining.
Applying these new results for entropic relations, we arrive at the
HUR that restricts measurements for any value of coarse graining.

\section{Uncertainty relations associated with continuous probability distributions of position and momentum variables}
\label{sec:introduction} 
For a quantum state described by a density operator $\hat{\varrho}$,
the probability densities describing measurements of $\hat{x}$ and $\hat{p}$
are given by 
\begin{equation}
\rho(x)=\langle x|\hat{\varrho}|x\rangle\;\;\mbox{and}\;\;\tilde{\rho}(p)=\langle p|\hat{\varrho}|p\rangle.
\label{disttrue}
\end{equation}
Since $\rho(x)$ and $\tilde{\rho}(p)$ are probability distributions corresponding to a quantum state, they obey uncertainty relations such as the HUR \eqref{Heisenberg-uncertainty-rel}, which involves variances that can be calculated from continuous probability distributions $\rho(x)$ and $\tilde{\rho}(p)$ as
\begin{equation}
\sigma_{z}^{2}\left[f\right]\equiv\int_{\mathbb{R}}dz\, z^{2}f\left(z\right)-\left(\int_{\mathbb{R}}dz\, z\, f\left(z\right)\right)^{2},
\label{def-variance}
\end{equation}
where $f=\rho,\,\tilde\rho$ and $z=x,\,p$.
Probability densities $\rho(x)$ and $\tilde\rho(p)$ will also obey the uncertainty relation
for the continuous R{\'e}nyi entropies ($1/\alpha+1/\beta=2$, $\beta\geq1$)
\cite{bialynicki06,zozor07}, 
\begin{equation}
h_{\alpha}[\rho]+h_{\beta}[\tilde{\rho}]\geq-\frac{1}{2\left(1-\alpha\right)}\ln\left(\frac{\alpha}{\pi\hbar}\right)-\frac{1}{2\left(1-\beta\right)}\ln\left(\frac{\beta}{\pi\hbar}\right),
\label{old-Bialinicki-result}
\end{equation}
where the continuous R{\'e}nyi entropy  is defined as \cite{renyi61}: 
\begin{equation}
h_{\lambda}[f]\equiv\frac{1}{1-\lambda}\ln\left(\int_{\mathbb{R}}dz\,\left[f\left(z\right)\right]^{\lambda}\right).\label{def-Shannon-entropy}
\end{equation}
The limit $\lambda\rightarrow1$ corresponds to the continuous Shannon
entropy 
\begin{equation}
\lim_{\lambda\rightarrow1}h_{\lambda}[f]=h\left[f\right]\equiv-\int_{\mathbb{R}}\; dz\, f\left(z\right)\;\ln\left(f(z)\right),
\end{equation}
and  so the entropic uncertainty relation for Shannon entropies is \cite{bialynicki75,bialynicki84}:
\begin{equation}
h[\rho]+h[\tilde{\rho}]\geq \ln \pi e \hbar.
\label{eq:shannon}
\end{equation}
Note that the uncertainty relations 
(\ref{Heisenberg-uncertainty-rel}), (\ref{old-Bialinicki-result}) and 
(\ref{eq:shannon}) involve the perfect knowledge of the continuous probability distributions $\rho(x)$ and $\tilde{\rho}(p)$ that can only be obtained from measurements with infinite precision.

\section{Uncertainty relations and coarse-grained measurements of position and momentum}

\label{sec:probability}

 In general, measurements are performed with finite precision, so 
any uncertainty relation involving infinite precision quantities  has to be adapted  
for  experimentally obtained quantities that depend on the coarse-grained nature of 
the measurement.  The usual uncertainty relations should be recovered from the coarse-grained uncertainty relations in the limit of infinite precision.
This leads to the interesting question:
what is the minimum precision (or maximum coarse graining) that still allows for a legitimate uncertainty relation?  Or equivalently: what is the minimum number of measurements, whose results belong to a fixed range of eigenvalues of the complementary observables $x$ and $p$, that allows one to verify an uncertainty relation?
The coarse-grained measurement process is equivalent to considering the probability distributions $\rho(x)$ and $\tilde{\rho}(p)$  sampled in bins, with
finite width $\Delta$ and $\delta$ for position and momentum, respectively.
These parameters shall coincide with the finite widths of the detectors used.
Due to these finite widths, the operators that are in fact measured are the coarse-grained
position and momentum operators, define as \cite{Rudnicki2011}:
\begin{equation}
\hat{x}_{\Delta}=\sum_{k}x_{k}\int_{(k-1/2)\Delta}^{(k+1/2)\Delta}dx\,\left|x\right\rangle \left\langle x\right|,\label{pcoar}
\end{equation}
and 
\begin{equation}
\hat{p}_{\delta}=\sum_{l}p_{l}\int_{(l-1/2)\delta}^{(l+1/2)\delta}dp\,\left|p\right\rangle \left\langle p\right|,\label{mcoar}
\end{equation}
where $x_{k}=k\Delta$ and $p_{l}=l\delta$ are the coordinates at
the center of the sampling windows. From repeated measurements over
identically prepared systems, we can construct the probabilities $r_{k}^{\Delta}$
and $s_{l}^{\delta}$ to obtain the results $x_{k}$ and $p_{l}$,
respectively. If the position and momentum measurements
are repeated with sufficient statistics, these probabilities
are expected to be very close to the actual values.  
Since $\left\langle \hat{x}_{\Delta}\right\rangle =\textrm{Tr}\left(\hat{\varrho}\cdot\hat{x}_{\Delta}\right)$
and $\left\langle \hat{p}_{\delta}\right\rangle =\textrm{Tr}\left(\hat{\varrho}\cdot\hat{p}_{\delta}\right)$, we obtain
\begin{equation}
\left\langle \hat{x}_{\Delta}\right\rangle =\sum_{k}x_{k}r_{k}^{\Delta},\qquad\left\langle \hat{p}_{\delta}\right\rangle =\sum_{l}p_{l}s_{l}^{\delta},
\end{equation}
where
\begin{equation}
r_{k}^{\Delta}=\int_{\left(k-1/2\right)\Delta}^{\left(k+1/2\right)\Delta}dx\,\rho\left(x\right),\quad s_{l}^{\delta}=\int_{\left(l-1/2\right)\delta}^{\left(l+1/2\right)\delta}dp\,\tilde{\rho}\left(p\right).\label{rs}
\end{equation}
 \par
The discrete variances that correspond to the measurements of the coarse-grained position and momentum operators are:
\bea
\sigma_{x_{\Delta}}^{2}&\equiv&\left\langle \hat{x}_{\Delta}^2\right\rangle-
\left\langle \hat{x}_{\Delta}\right\rangle^2  =\nonumber\\
&=&\sum_{k}x_{k}^{2}r_{k}^{\Delta}-\left(\sum_{k}x_{k}r_{k}^{\Delta}\right)^{2},\label{def-discrete-position-variance}
\eea
 and 
\bea
\sigma_{p_{\delta}}^{2}&\equiv&\left\langle \hat{p}_{\delta}^2\right\rangle-
\left\langle \hat{p}_{\delta}\right\rangle^2=\nonumber\\
&=&\sum_{l}p_{l}^{2}s_{l}^{\delta}-\left(\sum_{l}p_{l}s_{l}^{\delta}\right)^{2}.
\label{def-discrete-momentum-variance}
\eea

In the case where the widths $\Delta$ and $\delta$
are sufficiently small, these discrete variances are approximations
of the continuous variances $\sigma_{x}^{2}\equiv\sigma_{x}^{2}\left[\rho\right]$
and $\sigma_{p}^{2}\equiv\sigma_{p}^{2}\left[\tilde{\rho}\right]$
in Eq.(\ref{Heisenberg-uncertainty-rel}):
\begin{equation}
\lim_{\Delta\rightarrow0}\;\sigma_{x_{\Delta}}^{2}=\sigma_{x}^{2},\qquad\lim_{\delta\rightarrow0}\;\sigma_{p_{\delta}}^{2}=\sigma_{p}^{2}\;.\label{heis_limit_oper}
\end{equation}
However, as the sampling widths increase, 
the inferred variances $\sigma_{x_{\Delta}}^{2}$ and $\sigma_{p_{\delta}}^{2}$
begin to underestimate the true variances $\sigma_{x}^{2}$ and $\sigma_{p}^{2}$. In fact, we have the limit 
\begin{equation}
\lim_{\Delta\rightarrow\infty}\;\sigma_{x_{\Delta}}^{2}=0=\lim_{\delta\rightarrow\infty}\;\sigma_{p_{\delta}}^{2}.
\label{eq:HURlim}
\end{equation}
One possible adaptation of the HUR (\ref{Heisenberg-uncertainty-rel})
to finite coarse-grained measurements could be the usual Heisenberg-type 
uncertainty relation associated with any non-commuting observables,
\beq
\sigma_{x_{\Delta}}^{2}\sigma_{p_{\delta}}^{2}\ge \frac{1}{4}\left|\langle\left[
\hat x_{\Delta},\hat p_{\delta}\right]\rangle\right|^2 .
\label{comnorm}
\eeq
Unfortunately, this lower bound depends
on the state of the quantum system and is not useful from an 
experimental point of view. Furthermore, one can show that due to the coarse graining ($\Delta \neq 0$ or $\delta \neq 0$) there are always families of localized states for which the right hand side of (\ref{comnorm}) becomes equal to $0$.
\par
Another way to obtain uncertainty relations associated with coarse-grained measurements is to  investigate the properties of the probability
distributions $\left\{ r_{k}^{\Delta}\right\} $ and $\left\{ s_{l}^{\delta}\right\} $
using the discrete R{\'e}nyi entropies \cite{renyi61}: 
\begin{equation}
H_{\alpha}[r_{k}^{\Delta}]=\frac{1}{1-\alpha}\ln\sum_{k=-\infty}^{\infty}\left(r_{k}^{\Delta}\right)^{\alpha},\label{discrete-entropies}
\end{equation}
\begin{equation}
H_{\beta}[s_{l}^{\delta}]=\frac{1}{1-\beta}\ln\sum_{l=-\infty}^{\infty}\left(s_{l}^{\delta}\right)^{\beta}.\label{discrete-entropies2}
\end{equation}
In the limit $\alpha\rightarrow1$, $\beta\rightarrow1$ these definitions
recover the usual Shannon entropies $H[r_{k}^{\Delta}]=\lim_{\alpha\rightarrow1}H_{\alpha}[r_{k}^{\Delta}]=-\sum_k r_k^{\Delta}\ln r_k^{\Delta}$,
$H[s_{l}^{\delta}]=\lim_{\beta\rightarrow1}H_{\beta}[s_{l}^{\delta}]=
-\sum_l s_l^{\delta}\ln s_l^{\delta}$.
As in the case of the variances above, the discrete R\'enyi entropy 
starts to underestimate the continuous entropy when the widths 
$\Delta$ and $\delta$ are large.
In fact, when the sampling widths are extremely large, we have one 
$r_k^\Delta$ and one $s_l^\delta$ with near unit probabilities, which are responsible for the zero uncertainty in the discrete variables, 
{\it i.e.}
\begin{equation}
\lim_{\Delta\rightarrow\infty}\;H_{\alpha}[r_{k}^{\Delta}]=0=\lim_{\delta\rightarrow\infty}\;H_{\beta}[s_{l}^{\delta}].
\label{eq:renyilim}
\end{equation}
But in this case we have also the opposite situation, {\it i.e.} when the coarse-grained measurement is  fine,
$H[r_k^{\Delta}]$ and $H[s_l^{\delta}]$ start to super-estimate the Shannon entropies $h[\rho]$ and $h[\tilde\rho]$ respectively. 
In fact,  we have the limit situation  
$\lim_{\Delta\rightarrow 0}\lim_{\delta \rightarrow 0}
(H[r_k^{\Delta}]+H[s_l^{\delta}])=\infty$ 
\footnote{Note that for a discrete entropy  $H[f_j^{\eta}]$, where 
$f_j^{\eta}\equiv \int_{(j-1/2)\eta}^{(j+1/2)\eta}dz\; f(z)$,
we have in the limit  $\lim_{\eta\rightarrow 0} \left(H[f_j^{\eta}]+\ln(\eta)\right) \rightarrow h[f] $ \cite{cover}.}.
A first attempt to establish an uncertainty relation for coarse-grained 
measurement involving the discrete R\'enyi entropies, 
$H_{\alpha}[r_{k}^{\Delta}]$ and  
$H_{\beta}[s_{l}^{\delta}]$, was done by  Bialynicki-Birula \cite{bialynicki06}.
In section \ref{sec:entropic}, we will derive an improved lower bound for this uncertainty relation.
\par
In this paper,  we will be concerned with sampling widths that 
could be extremely large.   In the next section we will present the basic ingredients 
in order to obtain reliable uncertainty relations for the coarse-grained measurements.

\section{Continuous coarse-grained probability distribution functions for position and momentum}
\label{sec:four}
In calculations based on experimental data, we will show that it is advantageous to adopt the following approximated probability density functions (PDFs): 
\begin{equation}
w_{\Delta}\left(x\right)=\sum_{k=-\infty}^{\infty}r_{k}^{\Delta}D_{\Delta}\left(x,x_{k}\right),\label{def-w-Delta}
\end{equation}
and 
\begin{equation}
\tilde{w}_{\delta}\left(p\right)=\sum_{l=-\infty}^{\infty}s_{l}^{\delta}D_{\delta}\left(p,p_{l}\right).\label{def-tilde-w-delta}
\end{equation}
The $D_{\Delta}\left(x,x_{k}\right)$ and $D_{\delta}\left(p,p_{l}\right)$,
which we shall call generalized histogram functions (GHFs), are two independent
\textit{approximation to identity functions} \cite{Rudin1991} with width parameters $\Delta$ and $\delta$, respectively.   These are normalized functions: 
\begin{equation}
\int dz\; D_{\eta}(z,z_{j})=1, 
\label{eq:Dnorm}
\end{equation}
that converge
to the Dirac delta function: $\lim_{\eta\rightarrow0}=D_{\eta}(z,z_{j})=\delta(z-z_{j})$.  In this limit, we have
$\lim_{\Delta\rightarrow0}w_{\Delta}(x)=\rho(x)$ and $\lim_{\delta\rightarrow0}\tilde{w}_{\delta}(p)=\tilde{\rho}(p)$.
Additionally, we require that  $D_{\Delta}\left(x,x_{k}\right)$ and $D_{\delta}\left(p,p_{l}\right)$ have finite support
on the intervals $[(k-1/2)\Delta,(k+1/2)\Delta]$ and $[(l-1/2)\delta,(l+1/2)\delta]$,
respectively. 
We also impose the restriction 
that the bins are centered at the same points $z_j$ 
as the functions $D_{\eta}(z,z_{j})$, so we can recover these values through
\begin{equation}
z_{j}=\int_{\mathbb{R}}dz\, z\, D_{\eta}\left(z,z_{j}\right)=\int_{\left(j-1/2\right)\eta}^{\left(j+1/2\right)\eta}dz\, z\, D_{\eta}\left(z,z_{j}\right). \label{propAv}
\end{equation}
Moreover, we assume that these functions are translationally invariant:
\begin{equation}
D_{\eta}(z+z_{j}-z_{m},z_{j})=D_{\eta}(z,z_{m}),\label{propD}
\end{equation}
and we place no restriction on the variance of the function $D_{\eta}(z,z_{j})$. 
\par
\begin{figure}
\includegraphics[width=9cm]{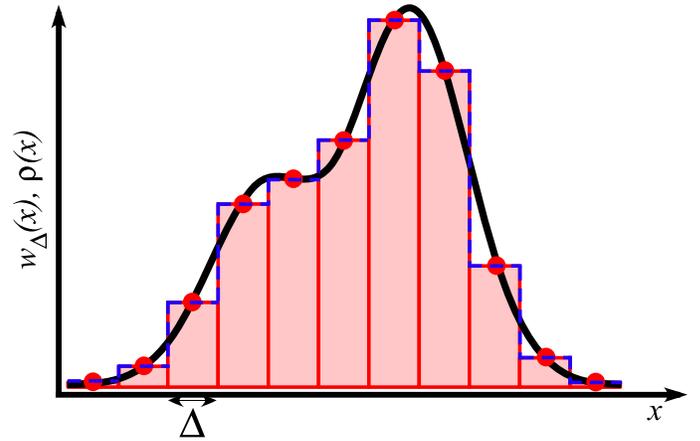}
\caption{(color online). Example of a continuous $w_{\Delta}(x)$ distribution function (blue dashed line) approximating the original distribution function $\rho(x)$ (black line) constructed out of rectangle functions.}
\label{fig:bin}
\end{figure}

Perhaps the simplest example of a GHF with these properties is the normalized rectangle
function: 
\begin{equation}
\mathrm{Rect}_{\eta}\left(z,z_{j}\right)=\begin{cases}
1/\eta & \textrm{ for }z\in\left[\left(j-\frac{1}{2}\right)\eta,\left(j+\frac{1}{2}\right)\eta\right]\\
0 & \textrm{ elsewhere}
\end{cases}\;.\label{Deta}
\end{equation}
An example of a $w_\Delta(x)$ distribution function constructed with rectangle functions is illustrated in Fig. \ref{fig:bin}.  The PDFs $w_\Delta(x)$ and $\tilde{w}_\delta(p)$ represent approximations to the actual PDFs $\rho(x)$ and $\tilde{\rho}(p)$ that are based on the results of the discretely-sampled measurements.   
\par
The advantage given by these new PDFs and the introduction of the histogram functions, is that they will allow us to derive reliable and optimal uncertainty relations
for coarse-grained variances (\ref{def-discrete-position-variance})
and (\ref{def-discrete-momentum-variance}).  
In the first step we shall show that there exists a simple relation between the
uncertainty associated with the approximate distribution--say--
$w_{\Delta}(x)$, the  
uncertainty of the associated discrete distribution $r^{\Delta}_k$, and
the uncertainty of the GHF $D_\Delta(x,x_k)$.   Let us show this connection only for the position variable since the calculations for the momentum variable are analogous. 
\par  
The variance of $w_{\Delta}(x)$ is 
\begin{align}
\sigma^2_{x}[w_{\Delta}] & \equiv
\int_{\mathbb{R}}dx\;x^2w_{\Delta}(x)-
\left(\int_{\mathbb{R}}dx\;x\;w_{\Delta}(x)\right)^2 \nonumber \\
& =
\sum_k\;r_k^{\Delta}\langle x^2\rangle_{k,D}-
\left(\sum_k\;r_k^{\Delta}\langle x\rangle_{k,D}\right)^2,
\label{eq:wvar}
\end{align}
where: 
\beq
\langle x^2\rangle_{k,D}\equiv\int_{(k-1/2)\Delta}^{(k+1/2)\Delta}\;dx\;x^2  D(x,x_k,\Delta),
\eeq
and
\beq
\langle x\rangle_{k,D}\equiv\int_{(k-1/2)\Delta}^{(k+1/2)\Delta}\;dx\;x\;  D(x,x_k,\Delta).  
\eeq
The variance of the function $D_\Delta(x,x_k)$ is then:
\beq
\sigma_\Delta^2\equiv \langle x^2\rangle_{k,D}-\langle x\rangle_{k,D}^2.
\eeq
Note that, due to the translational invariance of the GHF function (\ref{propD}), $\sigma_\Delta^2$,  in fact, does not depend on $k$ and we can write:
\beq
\label{sigmaD}
\sigma_\Delta^2=\sum_k\;r_k^{\Delta}\;\langle x^2\rangle_{k,D}-
\sum_k\;r_k^{\Delta}\langle x\rangle_{k,D}^2.
\eeq
Using Eq. (\ref{sigmaD}) to work out the total variance $\sigma^2_{x}[w_{\Delta}]$ we find that,
\beq
\sigma^2_{x}[w_{\Delta}]=\sigma_\Delta^2+
\sum_k\;r_k^{\Delta}\langle x\rangle_{k,D}^2-
\left(\sum_k\;r_k^{\Delta}\langle x\rangle_{k,D}\right)^2.
\eeq
Taking into account the property (\ref{propAv}) of the GHF we substitute $\langle x\rangle_{k,D}=x_k$ and finally obtain
\beq
\sigma^2_{x}[w_{\Delta}]=\sigma_{x_{\Delta}}^2+\sigma_\Delta^2,
\label{rel-fab}
\eeq 
 where the variance $\sigma_{x_{\Delta}}^2$ was defined in 
 Eq. \eqref{def-discrete-position-variance}. 
{ From Eq.  \eqref{rel-fab} it is easy to understand the limits
in Eq. \eqref{eq:HURlim}  if we interpret the two contributions to the variance
in Eq.(\ref{rel-fab}) (and in an analogous expression for the momentum) 
in the following way.  
First, we set the phase space origin 
at the center of the bins that contain $\langle \hat x_{\Delta}\rangle$ and $\langle \hat p_{\delta}\rangle$ ({\it i.e.} the central bins).  Thus, the first contribution in
Eq.  \eqref{rel-fab} is given by the discrete variances $\sigma^2_{x_{\Delta}}$ corresponding to the coarse-grained measurements 
$x_k=k\Delta$ outside the central bin, since the central bin has no contribution to the discrete variance in this case.  The other contribution,  $\sigma^2_{\Delta}$,   can be interpreted as the  variance of the GHF  of the central bin. 
For increasing values of coarse graining, the contribution to $\sigma^2_{x}[w_{\Delta}]$ from the central bin grows and the contribution from discrete measurements outside the central bin decreases.} 
 
For the uncertainty quantified by the continuous Shannon entropy, we have
\begin{align}
h[\omega_{\Delta}] & \equiv - \int_{\mathbb{R}}dx\; w_{\Delta}(x)\;\ln[w_{\Delta}(x)] 
\nonumber \\
& =
-\sum_k\;\int_k\;dx\;r_k^{\Delta} D_\Delta(x,x_k) \ln\left[
r_k^{\Delta} D_\Delta(x,x_k)\right] \nonumber \\
&= H[r_k^{\Delta}]+\sum_k\;r_k^{\Delta}\;h[D_\Delta(x,x_k)]  \nonumber \\
& =H[r_k^{\Delta}]+h_\Delta, \label{rel-entropy0}
\end{align}
{where the second line follows from the fact that the GHF has a compact support on the interval, and only one term inside the logarithm survives.}  Here $h_{\Delta}\equiv h\left[D_{\Delta}(x,x_k)\right]$ is the Shannon entropy of the continuous probability  distribution
$D(x,x_k,\Delta)$, which, according to (\ref{propD}), also does not depend on the index $k$.   As mentioned above,  similar results are found for the momentum distribution: 
\begin{equation}
\sigma_{p}^{2}\left[\tilde{w}_{\delta}\right]=\sigma_{p_{\delta}}^{2}+\sigma_{\delta}^{2},\label{relVariance}
\end{equation}
\begin{equation}
 h\left[\tilde{w}_{\delta}\right]=H\left[s_{l}^{\delta}\right]+h_{\delta}.\label{rel-entropy}
\end{equation}
It is important to realize that both $\sigma_{\eta}^{2}$ and $h_{\eta}$ do not
depend on the specific value $z_j$ of the center of each bin, so the uncertainty measured by these quantities
is associated with {\it a generic bin of the experimental sampling}.  
Thus, the variance (Shannon entropy) of the approximated PDFs are given by the sum of the discrete variance (Shannon entropy) of the  experimental points and the GHFs used.  This important property will allow us to construct consistent uncertainty relations in the next sections.  

\section{Entropic uncertainty relations for coarse-grained observables}
\label{sec:entropic}
\begin{figure}
\includegraphics[scale=0.82]{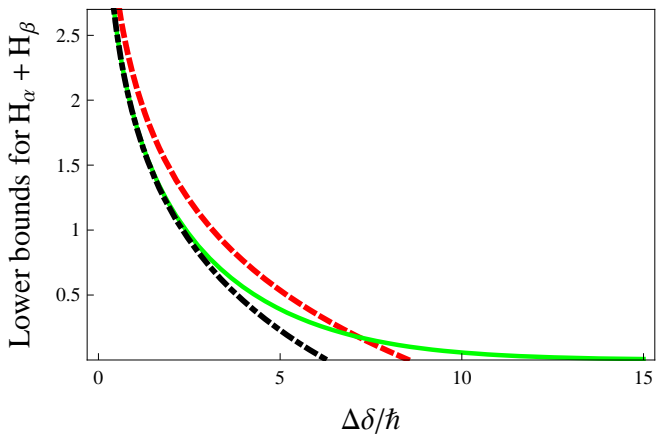}
\caption{(color online). We plot the lower bounds $\mathcal{R}$ (green/full), $\mathcal{B}_{1}$
(red/dashed) and $\mathcal{B}_{1/2}$ (black/dashed-dotted). 
Note that for $\alpha=1/2$ we have $L_{1/2}=\mathcal{R}$.
The lower bound $L_{\alpha}$ is optimal for $\alpha=1/2$.
}
\label{Fig1}
\end{figure}

We will first derive new uncertainty relations for coarse-grained measurements based on the R{\'e}nyi entropy.  
It was shown by Bialynicki-Birula that the discrete R\'enyi entropies (\ref{discrete-entropies})
and (\ref{discrete-entropies2}) satisfy the following entropic uncertainty
relation \cite{bialynicki06} ($1/\alpha+1/\beta=2$, $\beta\geq1$):
\begin{equation}
H_{\alpha}[r_{k}^{\Delta}]+H_{\beta}[s_{l}^{\delta}]\geq\mathcal{B}_{\alpha},
\label{IBB}
\end{equation}
where 
\begin{equation}
\mathcal{B}_{\alpha} = 
-\frac{1}{2}\left(\frac{\ln\alpha}{1-\alpha}+\frac{\ln\beta}{1-\beta}\right)-\ln\left(\frac{\Delta\delta}{\pi\hbar}\right).
\end{equation}
We note a full symmetry between the parameters $\alpha$ and $\beta$,
and since $\beta=\alpha/\left(2\alpha-1\right)$, we shall treat the
lower bound $\mathcal{B}_{\alpha}$ and further results as $\alpha$-dependent
increasing functions ($1/2\leq\alpha\leq1$):
\begin{equation}
-\ln\left(\frac{\Delta\delta}{2\pi\hbar}\right)=\mathcal{B}_{1/2}\leq\mathcal{B}_{\alpha}\leq\mathcal{B}_{1}=-\ln\left(\frac{\Delta\delta}{\pi e\hbar}\right).\label{limitation}
\end{equation}
Here we prove a new uncertainty
relation conjectured in \cite{rudnicki10}:
\begin{equation}
H_{\alpha}\left[r_{k}^{\Delta}\right]+H_{\beta}\left[s_{l}^{\delta}\right]\!\geq\!-\!\ln\left[\!\frac{\Delta\delta}{2\pi\hbar}\left[\! R_{00}\left(\frac{\Delta\delta}{4\hbar},1\right)\!\right]^{2}\right]\equiv\mathcal{R},\label{LR}
\end{equation}
where $R_{00}\left(\xi,\eta\right)$ %
\footnote{in the Wolfram Mathematica's notation it reads $\mathtt{SpheroidalS1[0,0,\xi,\eta]}$.%
} denotes one of the radial prolate spheroidal wave functions of the
first kind \cite{abramowitz}.  The full proof of this new relation is given in Appendix \ref{ap1}.  Since this relation is valid independently
of (\ref{IBB}), an improved lower bound for the sum of the R{\'e}nyi
entropies (\ref{discrete-entropies}) and (\ref{discrete-entropies2}) reads
\begin{equation}
H_{\alpha}\left[r_{k}^{\Delta}\right]+H_{\beta}\left[s_{l}^{\delta}\right]\!\geq\! L_{\alpha}\geq0,\label{optimal}
\end{equation}
where $L_{\alpha}\equiv\max\left\{\mathcal{B}_{\alpha},\mathcal{R}\right\}$.
When $\Delta\delta\ll\hbar$ we have 
\begin{equation}
\mathcal{R}\approx-\ln\left(\frac{\Delta\delta}{2\pi\hbar}\right)=\mathcal{B}_{1/2},
\end{equation}
so the final lower bound $L_{1/2}$ is a smooth function of $\Delta\delta/\hbar$.
For other values
of $\alpha$, especially $\alpha=1$, the $L_{\alpha}$ vs. $\Delta\delta/\hbar$ 
curve is not smooth.
 Figure \ref{Fig1} shows a plot of $\mathcal{R}$, $\mathcal{B}_{1/2}$
 and $\mathcal{B}_1$ as functions of $\Delta\delta/\hbar$. 
 Note that for $\alpha=1$ and 
 $\Delta\delta/\hbar\gtrsim 7$, we improve the lower bound $\mathcal{B}_1$ 
 in Eq.(\ref{IBB}) by  $L_1= \mathcal{R}$. We also note that for all values of $\alpha$ we have a non-trivial uncertainty relation since, unlike $\mathcal{B}_1$ and
 $\mathcal{B}_{1/2}$,  
 $L_{\alpha} > 0$ for $\Delta \delta < \infty$ (see Figure \ref{Fig1}). 
\section{Heisenberg uncertainty relations for coarse-grained observables}
\label{sec:Heisenberg}
The improved uncertainty relation \eqref{optimal} for the case of the Shannon entropy ($\alpha=1$) will allow us to derive a Heisenberg-like uncertainty relation for variances that is valid for any amount of coarse graining.  We first apply the reversed logarithmic
Sobolev inequality \cite{chafai} to the approximated PDFs  (\ref{def-w-Delta}) and (\ref{def-tilde-w-delta}):
\begin{equation}
\frac{1}{2}\ln\left(2\pi e\,\sigma_{x}^{2}\left[w_{\Delta}\right]\right)\geq h\left[w_{\Delta}\right],\label{sobolx}
\end{equation}
\begin{equation}
\frac{1}{2}\ln\left(2\pi e\,\sigma_{p}^{2}\left[\tilde{w}_{\delta}\right]\right)\geq h\left[\tilde{w}_{\delta}\right].\label{sobolp}
\end{equation}
Now we shall add these two inequalities, make use of relations (\ref{rel-entropy0}) and
(\ref{rel-entropy}) for the continuous entropies, and apply
the entropic uncertainty relation (\ref{optimal}) for $\alpha=\beta=1$.
We find that 
\bea
\sigma_{x}^{2}\left[w_{\Delta}\right]\sigma_{p}^{2}\left[\tilde{w}_{\delta}\right]
&\geq&\frac{\exp\left(2L_{1}\right)}{\left(2\pi e\right)^{2}}e^{2h_{\Delta}+2h_{\delta}}
\nonumber \\
&=&\frac{\hbar^2}{4}\frac{e^{2h_{\Delta}+2h_{\delta}}}{\Delta^2\delta^2}
g\left(\frac{\Delta\delta}{\hbar}\right)
\label{general}
\eea
where 
\beq
g\left(\frac{\Delta\delta}{\hbar}\right)\equiv\max\left\{1,\left(\frac{2}{e}\right)^2 \left[R_{00}\left(\frac{\Delta\delta}{4\hbar},1\right)\right]^{-4}
\right\}.
\eeq
In the case when the GHFs are the normalized rectangle functions in Eq.(\ref{Deta})
($\eta=\Delta,\delta$) we have the following Heisenberg-like  uncertainty relation, 
\beq
\sigma_{x}^{2}\left[w_{\Delta}\right]\sigma_{p}^{2}\left[\tilde{w}_{\delta}\right]
\geq
\frac{\hbar^2}{4}
g\left(\frac{\Delta\delta}{\hbar}\right).
\label{HUR-prl}
\eeq
When $\Delta\delta/\hbar < 6$ we have $g(\Delta\delta/\hbar)=1$ thus, this
uncertainty relation coincides with the result presented recently in
\cite{Rudnicki2011}:
\beq
\left(\sigma_{x_{\Delta}}^2+\frac{\Delta^2}{12}\right)
\left(\sigma_{p_{\delta}}^2+\frac{\delta^2}{12}\right)
\geq
\frac{\hbar^2}{4}.
\label{HUR-prl1}
\eeq 
Note that the uncertainty relation (\ref{HUR-prl1}) is satisfied trivially when $\Delta\delta/\hbar \geq 6$, so the uncertainty relation (\ref{HUR-prl}) seems to be an improvement of Eq.(\ref{HUR-prl1}) when  $\Delta\delta/\hbar \geq 6$. However,  we have to realize that both sides of (\ref{HUR-prl}) grow with coarse graining and the lower bound in
(\ref{HUR-prl}) grows slower that the left-hand side. As a result, the uncertainty relation (\ref{HUR-prl}) is also trivially satisfied for $\Delta\delta/\hbar \geq 6$, there is no improvement with respect to the previous result.

Nevertheless, the right-hand side of the inequality in Eq.(\ref{general})
contains information about the 
GHFs that can be optimized, since the variances $\sigma_{x}^{2}\left[w_{\Delta}\right]$
and $\sigma_{p}^{2}\left[\tilde{w}_{\delta}\right]$ are inferred directly from measurements.
We can now ask, what choice of GHFs gives us the optimal uncertainty relation? To answer this question, we perform  an optimization
procedure over the possible functional forms of the 
functions $D_\eta$ and also on the values of their variances.  All details are presented in Appendix \ref{ap3}. 
The solution of this optimization
procedure is obtained  for a GHF given by
a Gaussian function whose  support is  in the interval $[-\eta/2,\eta/2]$:
\begin{equation}
D^{opt}_{\eta}\left(z,0\right)=\sqrt{\frac{a_{\eta}}{\pi}}\frac{e^{-a_{\eta}z^{2}}}{\textrm{Erf}\left(\eta\sqrt{a_{\eta}}/2\right)},\qquad a_{\eta}\in\mathbb{R},\label{opti1}
\end{equation}
where $\textrm{Erf}$
is the usual error function and $a_{\eta}$ is an optimization parameter related to the
variance $\sigma_\eta^2$.  
With this optimal GHF
we arrive at the optimal coarse-grained version of the Heisenberg
uncertainty relation (see Appendix \ref{ap3}):
\bea
K\left(\frac{\sigma_{x_{\Delta}}^{2}}{\Delta^{2}}\right)K\left(\frac{\sigma_{p_{\delta}}^{2}}{\delta^{2}}\right)
&\geq&
\exp\left(2L_{1}\right)\nonumber \\
&=&\left(\frac{\pi  e \hbar}{\Delta\delta}\right)^2
g\left(\frac{\Delta\delta}{\hbar}\right)
\label{coar-grain-Heis}
\eea
where
\begin{equation}
K\left(u\right)=\frac{\exp\left[2u\mathcal{M}^{-1}\left(u\right)\right]}{\textrm{Erf}^{2}\left(\sqrt{\mathcal{M}^{-1}\left(u\right)}/2\right)},\label{F}
\end{equation}
and $\mathcal{M}^{-1}\left(\cdot\right)$ denotes the inverse of the
following invertible  function $\mathcal{M}\left(\cdot\right)$:
\begin{equation}
\mathcal{M}\left(t\right)=\frac{\exp\left(-t/4\right)}{2\sqrt{\pi t}\,\textrm{Erf}\left(\sqrt{t}/2\right)}.\label{coar-end}
\end{equation}
\begin{figure}
\includegraphics[scale=0.82]{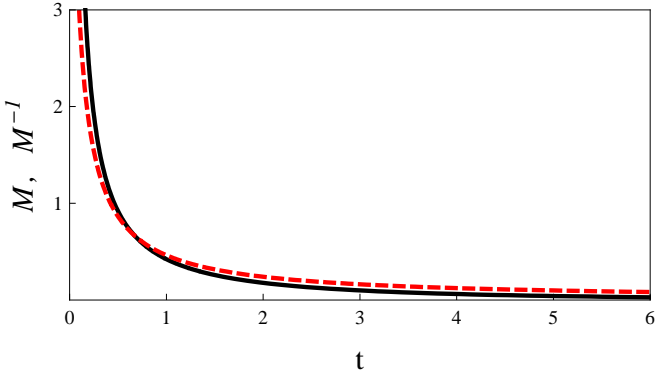}
\caption{(color online). Plots of the $\mathcal{M}\left(t\right)$ function
(black) and the $\mathcal{M}^{-1}\left(t\right)$ inverse function
(red/dashed).}
\label{Fig2}
\end{figure}
A plot of $\mathcal{M}$ and $\mathcal{M}^{-1}$ is shown in Fig. \ref{Fig2}.  In Fig. \ref{KfunctionPlot} we plot the function $K(u)$ in comparison 
to the linear function  $1+2\pi e\, u$. In the next section, we will analyze the new uncertainty relation \eqref{coar-grain-Heis}.  

\subsection{Analysis of the uncertainty relation Eq.(\ref{coar-grain-Heis})}

A first observation about the uncertainty relation (\ref{coar-grain-Heis}) 
is that it is valid for any finite value of coarse graining 
such that $\Delta\neq 0$ and 
$\delta\neq 0$, because in its 
derivation no restrictions were made on their possible values.
In the case of the limiting situation where  $\Delta,\delta\rightarrow0$ we 
recover the infinite precision Heisenberg uncertainty relation 
(\ref{Heisenberg-uncertainty-rel}).
In order to prove this we avoid the divergence on the right hand side of (\ref{coar-grain-Heis})
by multiplying both sides  by
the factor $\left(\Delta\delta\right)^{2}$ and calculate the limits $\Delta,\delta\rightarrow0$
in the following way ($g(0)=\max\left\{1,\left(2/e\right)^2 \right\}=1$):
\begin{equation}
\lim_{\Delta\rightarrow0}\left[\Delta^{2}K\left(\sigma_{x_{\Delta}}^{2}/\Delta^{2}\right)\right]\lim_{\delta\rightarrow0}\left[\delta^{2}K\left(\sigma_{p_{\delta}}^{2}/\delta^{2}\right)\right]\geq\left(\pi e\hbar\right)^{2}.\label{prelimit1}
\end{equation}
We can perform both limits in \eqref{prelimit1} separately. Let us now
introduce a new variable $v=\mathcal{M}^{-1}\left(\sigma_{x_{\Delta}}^{2}/\Delta^{2}\right)$.
Taking into account the limit (\ref{heis_limit_oper}), we obtain
\begin{align}
\lim_{\Delta\rightarrow0}\left[\Delta^{2}K\left(\sigma_{x_{\Delta}}^{2}/\Delta^{2}\right)\right]= & \,\sigma_{x}^{2}\lim_{v\rightarrow0}\left[\frac{1}{\mathcal{M}\left(v\right)}\frac{\exp\left[2v\mathcal{M}\left(v\right)\right]}{\textrm{Erf}^{2}\left(\sqrt{v}/2\right)}\right]\nonumber \\
 & =2\pi e\sigma_{x}^{2}.
\end{align}
\begin{figure}
\begin{centering}
\includegraphics[scale=0.82]{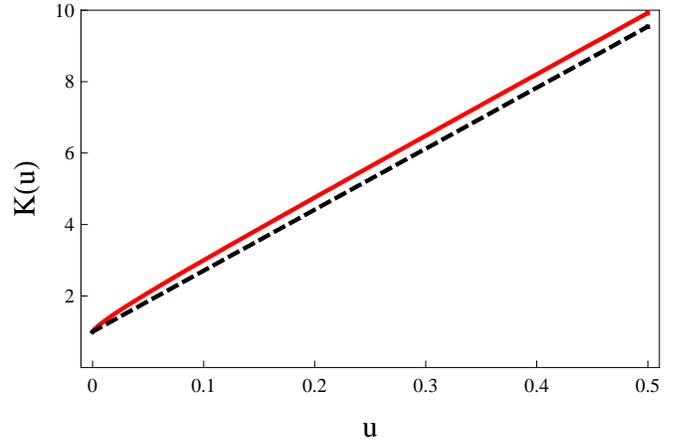}
\par\end{centering}
\caption{(color online). Comparison between the function {$K\left(u\right)$ (red) and the linear function $1+2\pi e\, u$ (black/dashed).}}
\label{KfunctionPlot}
\end{figure}
The same result can be obtained for the limit $\delta\rightarrow0$,
thus (\ref{prelimit1}) reads:
$\left(2\pi e\sigma_{x}\sigma_{p}\right)^{2}\geq\left(\pi e\hbar\right)^{2}$,
which is equivalent to the usual HUR in Eq.(\ref{Heisenberg-uncertainty-rel}).
\par
In the opposite limit of infinite coarse graining, the discrete variances go to $0$, as we mentioned in Eq.(\ref{eq:HURlim}) and discussed after  Eq.(\ref{rel-fab}).
Even for finite coarse graining, it is possible that  $\sigma_{x_{\Delta}}^{2}=0$ {\it or} $\sigma_{p_{\delta}}^{2}=0$, if the quantum state is localized in position or momentum (position or momentum probability distribution has a compact support).
However, the quantum state cannot be simultaneously localized in both position and momentum spaces. Therefore, 
for finite coarse graining it is forbidden that
$\sigma_{x_{\Delta}}^{2}=0=\sigma_{p_{\delta}}^{2}$.
Let us show that this fact is present in our uncertainty relation Eq.(\ref{coar-grain-Heis}). To this end, we shall calculate the limit $\sigma_{x_{\Delta}},\sigma_{p_{\delta}}\rightarrow0$.
Using the same variable $v$ as above, we have
\begin{equation}
\lim_{\sigma_{x_{\Delta}}\rightarrow0}K\left(\sigma_{x_{\Delta}}^{2}/\Delta^{2}\right)=\lim_{v\rightarrow\infty}\left[\frac{\exp\left[2v\mathcal{M}\left(v\right)\right]}{\textrm{Erf}^{2}\left(\sqrt{v}/2\right)}\right]=1.
\end{equation}
The same result can be obtained for the limit $\sigma_{p_{\delta}}\rightarrow0$.
Next, we note that for $\Delta<\infty$ and $\delta<\infty$ we have $L_{1}>0$.  Finally, when we put $\sigma_{x_{\Delta}}^{2}=0=\sigma_{p_{\delta}}^{2}$,
we obtain from (\ref{coar-grain-Heis}) the hierarchy of contradictory
inequalities:
\begin{equation}
1\geq\exp\left(2L_{1}\right)>1.\label{validity}
\end{equation}
Thus, zero variance in both discrete variables: $\sigma_{x_{\Delta}}^{2}=0=\sigma_{p_{\delta}}^{2}$, is prohibited.  

\begin{figure}
\includegraphics[scale=0.6]{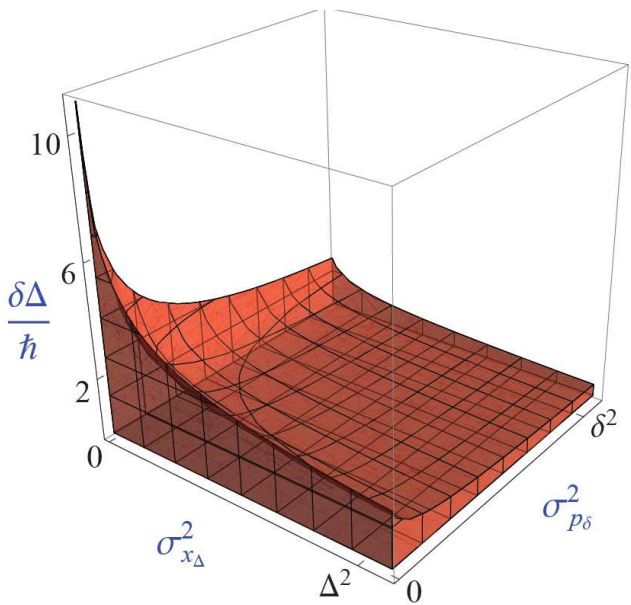}
\caption{(color online).  Plot of the uncertainty relation \eqref{coar-grain-Heis} as a function of the discrete variances $\sigma_{x_{\Delta}}^{2}$ and $\sigma_{p_{\delta}}^{2}$.  The red forbidden region shows that there is always an uncertainty relation for any size coarse graining.}
\label{Fig3}
\end{figure}

The coarse-grained Heisenberg uncertainty relation is shown graphically in
Figure \ref{Fig3}. The red area represents forbidden values of $\sigma_{x_{\Delta}}^{2}$
and $\sigma_{p_{\delta}}^{2}$. The narrow peak of forbidden values
for small $\sigma_{x_{\Delta}}^{2}$ and $\sigma_{p_{\delta}}^{2}$
illustrates the result (\ref{validity}).  Though this forbidden region gets smaller as $\Delta\delta/\hbar$ grows, there is always some forbidden region which limits the information that can be obtained about the non-commuting observables.  Thus, there always exists an uncertainty relation, regardless of the size of the coarse graining.  
{This is surprising since it is commonly argued  that in the large quantum number limit (semiclassical regime)  
we should recover classical mechanics for coarse-grained averaging due to finite-precision detectors \cite{ballentine98,kofler07}. We can follow the argumentation
in a simple example of a particle with mass $M$ in a one-dimensional infinite square well potential \cite{ballentine98}.  The energy eigenstates are  $\Psi_n(x) = \sqrt{2/L} \sin(n \pi x /L)$ ($0\leq x \leq L$). 
The eigenenergies are $E_n=p_n^2/2M$ and the two values of the momentum $p_n=\pm\hbar n\pi/L$
are equally probable since $\tilde \rho_n( p)=|\tilde\Psi_n( p)|^2$ consists of a symmetric 
probability distribution function peaked at $p_n$, with oscillatory tails that go to $\pm \infty$. 
It is easy to see that, for some finite coarse graining, we recover the classical probability distributions inside each bin, both in position and momentum, in the limit of large
quantum numbers, {\it i.e.} $n\rightarrow \infty$.
In the position representation we recover a constant value inside each bin.  In the momentum representation we obtain a zero value for each 
bin, except for the bins that contain the values
$p_n$, corresponding to Dirac delta functions that moves away to infinity. 
However, the limit $n\rightarrow \infty$ has only a formal meaning and, in fact, does not appear in real systems.
For large, but finite quantum number $n$, the position and momentum probability distributions are close, but not equal to the classical distributions. Thus, even if we choose coarse graining in the position representation that is  equal to the size $L$ of the potential well, so that $\sigma^2_{x_{\Delta}}=0$, and extremely large coarse graining 
in momentum, the value of $\sigma^2_{p_{\delta}}$ 
will contain contributions from the bins at the tails of the distribution
$\tilde \rho_n( p)$ that are different from zero for $-\infty < p < \infty$. 
In other words,  $\sigma^2_{p_{\delta}} \neq 0$, and in fact must be limited by the uncertainty relation in Eq.(\ref{coar-grain-Heis}).}
 
\section{Conclusions}
We have derived several new uncertainty relations for continuous variable quantum systems when coarse-grained measurements are performed.  First, we show a new bound for the uncertainty relation 
involving R\'enyi entropy.  This bound is an improvement for all entropy orders $\alpha$ for larger coarse graining, and is optimal for order $\alpha=1/2$.  Using this result, we derive a new Heisenberg-like uncertainty relation.  Surprisingly, there is always a meaningful uncertainty relation for any amount of finite coarse graining.  Thus, even coarse-grained measurements never commute, and information obtained about one observable increases uncertainty in the other. These results are interesting from a fundamental point of view, and also may find application in a quantum information scenario.  In particular, the security of several quantum key distribution schemes with continuous variables rely on uncertainty relations.       

\begin{acknowledgements} We acknowledge financial support from the Brazilian funding agencies
CNPq and FAPERJ. This work was performed as part of the Brazilian
Instituto Nacional de Ci{\^e}ncia e Tecnologia - Informa{\c c}{\~a}o Qu{\^a}ntica (INCT-IQ).
This research was also supported by the grant number N N202 174039 from the Polish Ministry
of Science and Higher Education for the years 2010\textendash{}2012. LR is specially indebted to Iwo Bialynicki-Birula who inspired and supported his efforts in the topic of entropic uncertainty relations.

\end{acknowledgements}

\appendix

\section{Derivation of the entropic uncertainty relation (\ref{LR})}

\label{ap1}To begin, let us consider a normalized, pure quantum
state described in position space by a one dimensional wave function
$\psi\left(x\right)$. The same state is described in momentum
space by $\tilde{\psi}\left(p\right)$ - the Fourier transform of
$\psi\left(x\right)$. These wave functions provide two probability
distributions:
\begin{equation}
g\left(x\right)=\left|\psi\left(x\right)\right|^{2},\qquad\tilde{g}\left(p\right)=\left|\tilde{\psi}\left(p\right)\right|^{2},\label{prob}
\end{equation}
in position and momentum space, respectively. For (\ref{prob})
we shall define, analogous to (\ref{rs}), the discrete probability
distributions:

\begin{equation}
q_{k}=\int_{\left(k-1/2\right)\Delta}^{\left(k+1/2\right)\Delta}dx\, g\left(x\right),\quad p_{l}=\int_{\left(l-1/2\right)\delta}^{\left(l+1/2\right)\delta}dp\,\tilde{g}\left(p\right).\label{rs-1}
\end{equation}

The starting point of our derivation shall be the definition of two
new probability distributions $\left\{ \left|a_{km}\right|^{2}\right\} $
and $\left\{ \left|b_{ln}\right|^{2}\right\} $ that are distinct from (\ref{rs-1}), where \cite{rudnicki10}:
\begin{equation}
a_{km}=\int_{\left(k-1/2\right)\Delta}^{\left(k+1/2\right)\Delta}\!\!\! dx\,\psi\left(x\right)\varphi_{km}^{*}\left(x\right),\label{proba}
\end{equation}
 
\begin{equation}
b_{ln}=\int_{\left(l-1/2\right)\delta}^{\left(l+1/2\right)\delta}\!\!\! dp\,\tilde{\psi}\left(p\right)\theta_{ln}^{*}\left(p\right).\label{probb}
\end{equation}
Functions $\varphi_{km}(x)$ have been arbitrarily chosen to form an orthonormal basis in
the $k$th bin:
\begin{equation}
\int_{\left(k-1/2\right)\Delta}^{\left(k+1/2\right)\Delta}\!\!\! dx\,\varphi_{km}\left(x\right)\varphi_{km'}^{*}\left(x\right)=\delta_{mm'}.\label{c2}
\end{equation}
Similarly, in momentum space, we introduce the
functions $\theta_{ln}(p)$, which are orthonormal in the in the $l$th bin:
\begin{equation}
\int_{\left(l-1/2\right)\delta}^{\left(l+1/2\right)\delta}\!\!\! dp\,\theta_{ln}\left(p\right)\theta_{ln'}^{*}\left(p\right)=\delta_{nn'}.\label{c3}
\end{equation}
For the probability distributions $\left\{ \left|a_{km}\right|^{2}\right\} $
and $\left\{ \left|b_{ln}\right|^{2}\right\} $, the Riesz theorem
\cite{riesz} reads ($1/\alpha+1/\beta=2,\quad\beta\geq1$):

\begin{equation}
\left[\mathcal{C}\sum_{n}\left|b_{ln}\right|^{2\beta}\right]^{1/\beta}\leq\left[\mathcal{C}\sum_{m}\left|a_{km}\right|^{2\alpha}\right]^{1/\alpha},\label{b8}
\end{equation}
where the constant $\mathcal{C}$ is:
\begin{equation}
\mathcal{C}=\!\!\!\!\sup_{(k,m,l,n)}\left|\int_{\left(k-1/2\right)\Delta}^{\left(k+1/2\right)\Delta}\!\!\! dx\int_{\left(l-1/2\right)\delta}^{\left(l+1/2\right)\delta}\!\!\! dp\,\frac{e^{ipx/\hbar}}{\sqrt{2\pi\hbar}}\varphi_{km}^{*}\left(x\right)\theta_{ln}\left(p\right)\right|.\label{b9}
\end{equation}

Using the well known Maassen-Uffink result \cite{maassen88} it was shown
that \cite{rudnicki10}:
\begin{equation}
H_{\alpha}\left[\left|a_{km}\right|^{2}\right]+H_{\beta}\left[\left|b_{ln}\right|^{2}\right]\geq-2\ln\mathcal{C},\label{c8}
\end{equation}
where the R{\'e}nyi entropies $H_{\alpha}\left[\left|a_{km}\right|^{2}\right]$
and $H_{\beta}\left[\left|b_{ln}\right|^{2}\right]$ are related to
the probability distributions $\left\{ \left|a_{km}\right|^{2}\right\} $
and $\left\{ \left|b_{ln}\right|^{2}\right\} $. To obtain this uncertainty
relation one needs to take the logarithm of both sides of (\ref{b8})
and recognize the definitions of the R{\'e}nyi entropies. In \cite{rudnicki10,schur}
it was also shown that:
\begin{equation}
\mathcal{C}<\exp\left(-\mathcal{R}/2\right)\equiv\sqrt{\frac{\Delta\delta}{2\pi\hbar}}R_{00}\left(\frac{\Delta\delta}{4\hbar},1\right).\label{c19}
\end{equation}
The integral equation leading to this result appears in the signal processing theory, and also in the topic of entropic uncertainty relations \cite{partovi83}.
Since the inequality (\ref{c19}) is independent of the choice of
the functions $\varphi_{km}(x)$ and $\theta_{ln}(p)$, we can take:
\begin{equation}
\varphi_{km}\left(x\right)=\begin{cases}
\psi\left(x\right)/\sqrt{q_{k}} & m=0\\
\textrm{orthogonal functions} & m\neq0
\end{cases},
\end{equation}
and:
\begin{equation}
\theta_{ln}\left(p\right)=\begin{cases}
\tilde{\psi}\left(p\right)/\sqrt{p_{l}} & n=0\\
\textrm{orthogonal functions} & n\neq0
\end{cases}.
\end{equation}
In this particular choice we have:
\begin{equation}
a_{km}=\delta_{0m}\sqrt{q_{k}},\quad b_{ln}=\delta_{0n}\sqrt{p_{l}},
\end{equation}
and:
\begin{equation}
H_{\alpha}\left[\left|a_{km}\right|^{2}\right]=H_{\alpha}\left[q_{k}\right],\qquad H_{\beta}\left[\left|b_{ln}\right|^{2}\right]=H_{\beta}\left[p_{l}\right].
\end{equation}
This observation, together with (\ref{c8}) and (\ref{c19}), leads to
the result $H_{\alpha}\left[q_{k}\right]+H_{\beta}\left[p_{l}\right]\geq\mathcal{R}$. 

In order to extend this uncertainty relation to the case of the probability
distributions (\ref{disttrue}) related to the mixed state density
operator $\hat{\varrho}$, we note that (\ref{disttrue}) can always be
represented in the following way ($\sum_{i}\lambda_{i}=1$):
\begin{equation}
\rho\left(x\right)=\sum_{i}\lambda_{i}g_{i}\left(x\right),\quad\textrm{ and }\quad\tilde{\rho}\left(p\right)=\sum_{i}\lambda_{i}\tilde{g}_{i}\left(p\right),\label{mixed}
\end{equation}
where $g_{i}\left(x\right)$ and $\tilde{g}_{i}\left(p\right)$ are
probability distributions of the form (\ref{prob}). Equation
(\ref{mixed}) immediately implies the same decomposition of the probability
distributions $\left\{ r_{k}^{\Delta}\right\} $ and $\left\{ s_{l}^{\delta}\right\} $:
\begin{equation}
r_{k}^{\Delta}=\sum_{i}\lambda_{i}q_{k}^{i},\quad\textrm{ and }\quad s_{l}^{\delta}=\sum_{i}\lambda_{i}p_{l}^{i},\label{mixed-1}
\end{equation}
where $q_{k}^{i}$ and $p_{l}^{i}$ are calculated for the probability
distributions $g_{i}\left(x\right)$ and $\tilde{g}_{i}\left(p\right)$
respectively. Using the arguments provided in \cite{bialynicki06,bialynicki11}, it follows from the Minkowski inequality that:
\begin{equation}
H_{\alpha}\left[r_{k}^{\Delta}\right] + H_{\beta}\left[s_{l}^{\delta}\right] \geq\sum_{i}\lambda_{i}\left ( H_{\alpha}\left[q_{k}^{i}\right]+H_{\beta}\left[p_{l}^{i}\right]\right ),
\end{equation}
which finishes the derivation.

\section{Optimization leading to Eqs. (\ref{coar-grain-Heis}-\ref{coar-end})}

\label{ap3}In the uncertainty relation (\ref{general}) there are
two pairs of parameters that are independent of the state $\hat{\varrho}$,\textit{
i.e.} $\left\{ h_{\Delta},\sigma_{\Delta}^{2}\right\} $ and $\left\{ h_{\delta},\sigma_{\delta}^{2}\right\} $.
We will perform an optimization over these parameters, but since
they are not independent among themselves we shall do this procedure in two steps.
First we maximize the Shannon entropies $h_{\Delta}$ and $h_{\delta}$,
keeping the variances $\sigma_{\Delta}^{2}$, $\sigma_{\delta}^{2}$
constant. We can consider the position and momentum variables separately,
thus we will perform all calculations for the general case of the
$D_{\eta}\left(z,0\right)$ function. To this end, we shall solve the
variational equation: 
\begin{align}
\frac{\delta}{\delta D_{\eta}\left(z,0\right)}\left(-\int_{-\eta/2}^{\eta/2}dz\, D_{\eta}\left(z,0\right)\ln\left(D_{\eta}\left(z,0\right)\right)\right.\\
\left.-\lambda_{\eta}\int_{-\eta/2}^{\eta/2}dz\, D_{\eta}\left(z,0\right)-a_{\eta}\int_{-\eta/2}^{\eta/2}dz\, z^{2}D_{\eta}\left(z,0\right)\right) & =0.\nonumber 
\end{align}
Here $\lambda_{\eta}$ and $a_{\eta}$ are $\eta$-dependent
Lagrange multipliers associated with the normalization and constant
variance constraints.
The solution is a Gaussian function with the support only in the interval $[-\eta/2,\eta/2]$
\begin{equation}
D_{\eta}\left(z,0\right)=\sqrt{\frac{a_{\eta}}{\pi}}\frac{e^{-a_{\eta}z^{2}}}{\textrm{Erf}\left(\eta\sqrt{a_{\eta}}/2\right)},\qquad a_{\eta}\in\mathbb{R},\label{opti1}
\end{equation}
where $\textrm{Erf}\left(y\right)=\frac{2}{\sqrt{\pi}}\int_{0}^{y}dt\,\exp\left(-t^{2}\right)$
denotes the usual error function. The normalization constraint gives
the value of $\lambda_{\eta}$ as a function of the $a_{\eta}$ parameter
\begin{equation}
\lambda_{\eta}\left(a_{\eta}\right)=\ln\left(\sqrt{\frac{\pi}{a_{\eta}}}\textrm{Erf}\left(\frac{\eta\sqrt{a_{\eta}}}{2}\right)\right)-1.
\end{equation}
The variance constraint imposes a relation between $a_{\eta}$
and the variance $\sigma_{\eta}^{2}$ of the following form
\begin{equation}
\sigma_{\eta}^{2}\left(a_{\eta}\right)=\frac{1}{2a_{\eta}}\left(1-\sqrt{\frac{a_{\eta}}{\pi}}\frac{\eta\exp\left(-a_{\eta}\eta^{2}/4\right)}{\textrm{Erf}\left(\eta\sqrt{a_{\eta}}/2\right)}\right).\label{parameter a}
\end{equation}
The right-hand side is a monotonically-decreasing invertible function
of the $a_{\eta}$ parameter.  Moreover, this relation restricts the
values of the variance to $0\leq\sigma_{\eta}^{2}<\eta^{2}/4$, what
is a natural consequence of the fact that the maximal value of $z^{2}$
on the interval $[-\eta/2,\eta/2]$ is equal to $\eta^{2}/4$. 
It is worth noticing that the case $\sigma_{\eta}^{2}=\eta^{2}/12$, which is
equivalent to $a_{\eta}=0$, describes the case of the rectangle
function (\ref{Deta}). From
now on, according to the relation (\ref{parameter a}), we shall use
the $a_{\eta}$ parameter instead of the variance $\sigma_{\eta}^{2}$. Due to the results (\ref{opti1}) and (\ref{parameter a})
we have
\begin{equation}
h_{\eta}\left(a_{\eta}\right)=1+\lambda_{\eta}\left(a_{\eta}\right)+a_{\eta}\sigma_{\eta}^{2}\left(a_{\eta}\right).\label{entropy a}
\end{equation}
This entropy attains its maximal value equal to $\ln\eta$ for $a_{\eta}=0$
(the rectangle function (\ref{Deta})).

Thus, we can rewrite the uncertainty relation (\ref{general}) optimized
with respect to both entropies in the following way:
\begin{equation}
\frac{\sigma_{x_{\Delta}}^{2}\!+\sigma_{\Delta}^{2}\left(a_{\Delta}\right)}{\exp\left(2h_{\Delta}\left(a_{\Delta}\right)\right)}\cdot\frac{\sigma_{p_{\delta}}^{2}\!+\sigma_{\delta}^{2}\left(a_{\delta}\right)}{\exp\left(2h_{\delta}\left(a_{\delta}\right)\right)}\!\geq\!\frac{\exp\left(2L_{1}\right)}{\left(2\pi e\right)^{2}}.
\end{equation}
Substituting results (\ref{parameter a}) and (\ref{entropy a}),
this relation reads 
\begin{equation}
F\left(\frac{\sigma_{x_{\Delta}}^{2}}{\Delta^{2}},\Delta^{2}a_{\Delta}\right)F\left(\frac{\sigma_{p_{\delta}}^{2}}{\delta^{2}},\delta^{2}a_{\delta}\right)\geq\exp\left(2L_{1}\right),\label{coar-grain-Heis-1}
\end{equation}
where:
\begin{equation}
F\left(u,t\right)=\frac{2t\left(u-\mathcal{M}\left(t\right)\right)+1}{\textrm{Erf}^{2}\left(\sqrt{t}/2\right)}\exp\left(2t\mathcal{M}\left(t\right)\right),\label{F_Sigma}
\end{equation}
and the $\mathcal{M}\left(\cdot\right)$ function has been defined
in (\ref{coar-end}). The second task in the optimization procedure
is to find the minimal value of $F\left(u,t\right)$ with respect
to $t\in\mathbb{R}$, where $u\geq0$ plays the role of an independent
parameter. Differentiation of (\ref{F_Sigma}) leads to the minimum
(the second derivative with respect to $t$ is a positive function
for $u\geq0$) at point $t_{min}=\mathcal{M}^{-1}\left(u\right)$.
Unfortunately there is no analytical expression for the $\mathcal{M}^{-1}\left(\cdot\right)$
function. In order to finish the derivation of the optimal uncertainty
relation (\ref{coar-grain-Heis}), we shall take the function (\ref{F})
to be $K\left(u\right)=F\left(u,\mathcal{M}^{-1}\left(u\right)\right)$.


\begin{thebibliography}{39}
\expandafter\ifx\csname natexlab\endcsname\relax\def\natexlab#1{#1}\fi
\expandafter\ifx\csname bibnamefont\endcsname\relax
  \def\bibnamefont#1{#1}\fi
\expandafter\ifx\csname bibfnamefont\endcsname\relax
  \def\bibfnamefont#1{#1}\fi
\expandafter\ifx\csname citenamefont\endcsname\relax
  \def\citenamefont#1{#1}\fi
\expandafter\ifx\csname url\endcsname\relax
  \def\url#1{\texttt{#1}}\fi
\expandafter\ifx\csname urlprefix\endcsname\relax\def\urlprefix{URL }\fi
\providecommand{\bibinfo}[2]{#2}
\providecommand{\eprint}[2][]{\url{#2}}

\bibitem[{\citenamefont{Reid}(2000)}]{reid00}
\bibinfo{author}{\bibfnamefont{M.~D.} \bibnamefont{Reid}},
  \bibinfo{journal}{Phys. Rev. A} \textbf{\bibinfo{volume}{62}},
  \bibinfo{pages}{062308} (\bibinfo{year}{2000}).

\bibitem[{\citenamefont{Grosshans and Cerf}(2004)}]{grosshans04}
\bibinfo{author}{\bibfnamefont{F.}~\bibnamefont{Grosshans}} \bibnamefont{and}
  \bibinfo{author}{\bibfnamefont{N.~J.} \bibnamefont{Cerf}},
  \bibinfo{journal}{Phys. Rev. Lett.} \textbf{\bibinfo{volume}{92}},
  \bibinfo{eid}{047905} (\bibinfo{year}{2004}).

\bibitem[{\citenamefont{Simon}(2000)}]{simon00}
\bibinfo{author}{\bibfnamefont{R.}~\bibnamefont{Simon}},
  \bibinfo{journal}{Phys. Rev. Lett.} \textbf{\bibinfo{volume}{84}},
  \bibinfo{pages}{2726} (\bibinfo{year}{2000}).

\bibitem[{\citenamefont{Duan et~al.}(2000)\citenamefont{Duan, Giedke, Cirac,
  and Zoller}}]{duan00}
\bibinfo{author}{\bibfnamefont{L.-M.} \bibnamefont{Duan}},
  \bibinfo{author}{\bibfnamefont{G.}~\bibnamefont{Giedke}},
  \bibinfo{author}{\bibfnamefont{J.~I.} \bibnamefont{Cirac}}, \bibnamefont{and}
  \bibinfo{author}{\bibfnamefont{P.}~\bibnamefont{Zoller}},
  \bibinfo{journal}{Phys. Rev. Lett.} \textbf{\bibinfo{volume}{84}},
  \bibinfo{pages}{2722} (\bibinfo{year}{2000}).

\bibitem[{\citenamefont{Mancini et~al.}(2002)\citenamefont{Mancini,
  Giovannetti, Vitali, and Tombesi}}]{mancini02}
\bibinfo{author}{\bibfnamefont{S.}~\bibnamefont{Mancini}},
  \bibinfo{author}{\bibfnamefont{V.}~\bibnamefont{Giovannetti}},
  \bibinfo{author}{\bibfnamefont{D.}~\bibnamefont{Vitali}}, \bibnamefont{and}
  \bibinfo{author}{\bibfnamefont{P.}~\bibnamefont{Tombesi}},
  \bibinfo{journal}{Phys. Rev. Lett.} \textbf{\bibinfo{volume}{88}},
  \bibinfo{eid}{120401} (\bibinfo{year}{2002}).

\bibitem[{\citenamefont{Nha and Zubairy}(2008)}]{nha08}
\bibinfo{author}{\bibfnamefont{H.}~\bibnamefont{Nha}} \bibnamefont{and}
  \bibinfo{author}{\bibfnamefont{M.~S.} \bibnamefont{Zubairy}},
  \bibinfo{journal}{Phys. Rev. Lett.} \textbf{\bibinfo{volume}{101}},
  \bibinfo{pages}{130402} (\bibinfo{year}{2008}).

\bibitem[{\citenamefont{Walborn et~al.}(2009)\citenamefont{Walborn, Taketani,
  Salles, Toscano, and de~Matos~Filho}}]{walborn09}
\bibinfo{author}{\bibfnamefont{S.~P.} \bibnamefont{Walborn}},
  \bibinfo{author}{\bibfnamefont{B.~G.} \bibnamefont{Taketani}},
  \bibinfo{author}{\bibfnamefont{A.}~\bibnamefont{Salles}},
  \bibinfo{author}{\bibfnamefont{F.}~\bibnamefont{Toscano}}, \bibnamefont{and}
  \bibinfo{author}{\bibfnamefont{R.~L.} \bibnamefont{de~Matos~Filho}},
  \bibinfo{journal}{Phys. Rev. Lett.} \textbf{\bibinfo{volume}{103}},
  \bibinfo{pages}{160505} (\bibinfo{year}{2009}).

\bibitem[{\citenamefont{Saboia et~al.}(2011)\citenamefont{Saboia, Toscano, and
  Walborn}}]{saboia11}
\bibinfo{author}{\bibfnamefont{A.}~\bibnamefont{Saboia}},
  \bibinfo{author}{\bibfnamefont{F.}~\bibnamefont{Toscano}}, \bibnamefont{and}
  \bibinfo{author}{\bibfnamefont{S.~P.} \bibnamefont{Walborn}},
  \bibinfo{journal}{Phys. Rev. A} \textbf{\bibinfo{volume}{83}},
  \bibinfo{pages}{032307} (\bibinfo{year}{2011}).

\bibitem[{\citenamefont{Reid}(1989)}]{reid89}
\bibinfo{author}{\bibfnamefont{M.~D.} \bibnamefont{Reid}},
  \bibinfo{journal}{Phys. Rev. A} \textbf{\bibinfo{volume}{40}},
  \bibinfo{pages}{913} (\bibinfo{year}{1989}).

\bibitem[{\citenamefont{Wiseman et~al.}(2007)\citenamefont{Wiseman, Jones, and
  Doherty}}]{wiseman07}
\bibinfo{author}{\bibfnamefont{H.~M.} \bibnamefont{Wiseman}},
  \bibinfo{author}{\bibfnamefont{S.~J.} \bibnamefont{Jones}}, \bibnamefont{and}
  \bibinfo{author}{\bibfnamefont{A.~C.} \bibnamefont{Doherty}},
  \bibinfo{journal}{Phys. Rev. Lett.} \textbf{\bibinfo{volume}{98}},
  \bibinfo{eid}{140402} (\bibinfo{year}{2007}).

\bibitem[{\citenamefont{Reid et~al.}(2010)\citenamefont{Reid, Drummond, Bowen,
  Cavalcanti, Lam, Bachor, Anderson, and Leuchs}}]{reid10}
\bibinfo{author}{\bibfnamefont{M.~D.} \bibnamefont{Reid}},
  \bibinfo{author}{\bibfnamefont{P.~D.} \bibnamefont{Drummond}},
  \bibinfo{author}{\bibfnamefont{W.~P.} \bibnamefont{Bowen}},
  \bibinfo{author}{\bibfnamefont{E.~G.} \bibnamefont{Cavalcanti}},
  \bibinfo{author}{\bibfnamefont{P.~K.} \bibnamefont{Lam}},
  \bibinfo{author}{\bibfnamefont{H.~A.} \bibnamefont{Bachor}},
  \bibinfo{author}{\bibfnamefont{U.~L.} \bibnamefont{Anderson}},
  \bibnamefont{and} \bibinfo{author}{\bibfnamefont{G.}~\bibnamefont{Leuchs}},
  \bibinfo{journal}{Rev. Mod. Phys.} \textbf{\bibinfo{volume}{81}},
  \bibinfo{pages}{1727} (\bibinfo{year}{2010}).

\bibitem[{\citenamefont{Walborn et~al.}(2011)\citenamefont{Walborn, Salles,
  Gomes, Toscano, and Souto~Ribeiro}}]{walborn11a}
\bibinfo{author}{\bibfnamefont{S.~P.} \bibnamefont{Walborn}},
  \bibinfo{author}{\bibfnamefont{A.}~\bibnamefont{Salles}},
  \bibinfo{author}{\bibfnamefont{R.~M.} \bibnamefont{Gomes}},
  \bibinfo{author}{\bibfnamefont{F.}~\bibnamefont{Toscano}}, \bibnamefont{and}
  \bibinfo{author}{\bibfnamefont{P.~H.} \bibnamefont{Souto~Ribeiro}},
  \bibinfo{journal}{Phys. Rev. Lett.} \textbf{\bibinfo{volume}{106}},
  \bibinfo{pages}{130402} (\bibinfo{year}{2011}).

\bibitem[{\citenamefont{Heisenberg}(1927)}]{heisenberg}
\bibinfo{author}{\bibfnamefont{W.}~\bibnamefont{Heisenberg}},
  \bibinfo{journal}{Z. Phys.} \textbf{\bibinfo{volume}{43}},
  \bibinfo{pages}{122} (\bibinfo{year}{1927}).
  
 { 
 \bibitem[{\citenamefont{Kennard}(1927)}]{kennard}
\bibinfo{author}{\bibfnamefont{E. H.}~\bibnamefont{Kennard}},
  \bibinfo{journal}{Z. Phys.} \textbf{\bibinfo{volume}{44}},
  \bibinfo{pages}{326} (\bibinfo{year}{1927}).
  
   \bibitem[{\citenamefont{Robertson}(1927)}]{robertson}
\bibinfo{author}{\bibfnamefont{H. P.}~\bibnamefont{Robertson}},
  \bibinfo{journal}{Phys. Rev.} \textbf{\bibinfo{volume}{34}},
  \bibinfo{pages}{163} (\bibinfo{year}{1929}).
 
  }


\bibitem[{\citenamefont{Bialynicki-Birula and Mycielski}(1975)}]{bialynicki75}
\bibinfo{author}{\bibfnamefont{I.}~\bibnamefont{Bialynicki-Birula}}
  \bibnamefont{and}
  \bibinfo{author}{\bibfnamefont{J.}~\bibnamefont{Mycielski}},
  \bibinfo{journal}{Commun. Math. Phys.} \textbf{\bibinfo{volume}{44}},
  \bibinfo{pages}{129} (\bibinfo{year}{1975}).

\bibitem[{\citenamefont{Deutsch}(1983)}]{deutsch83}
\bibinfo{author}{\bibfnamefont{D.}~\bibnamefont{Deutsch}},
  \bibinfo{journal}{Phys. Rev. Lett.} \textbf{\bibinfo{volume}{50}},
  \bibinfo{pages}{631} (\bibinfo{year}{1983}).


\bibitem[{\citenamefont{Partovi}(1983)}]{partovi83}
\bibinfo{author}{\bibfnamefont{M.~H.} \bibnamefont{Partovi}},
  \bibinfo{journal}{Phys. Rev. Lett.} \textbf{\bibinfo{volume}{50}},
  \bibinfo{pages}{1883} (\bibinfo{year}{1983}).

\bibitem[{\citenamefont{Bialynicki-Birula}(1984)}]{bialynicki84}
\bibinfo{author}{\bibfnamefont{I.}~\bibnamefont{Bialynicki-Birula}},
  \bibinfo{journal}{Phys. Lett.} \textbf{\bibinfo{volume}{103 A}},
  \bibinfo{pages}{253} (\bibinfo{year}{1984}).

\bibitem[{\citenamefont{Kraus}(1987)}]{kraus87}
\bibinfo{author}{\bibfnamefont{K.}~\bibnamefont{Kraus}},
  \bibinfo{journal}{Phys. Rev. D} \textbf{\bibinfo{volume}{35}},
  \bibinfo{pages}{3070} (\bibinfo{year}{1987}).

\bibitem[{\citenamefont{Maassen and Uffink}(1988)}]{maassen88}
\bibinfo{author}{\bibfnamefont{H.}~\bibnamefont{Maassen}} \bibnamefont{and}
  \bibinfo{author}{\bibfnamefont{J.~B.~M.} \bibnamefont{Uffink}},
  \bibinfo{journal}{Phys. Rev. Lett.} \textbf{\bibinfo{volume}{60}},
  \bibinfo{pages}{1103} (\bibinfo{year}{1988}).

\bibitem[{\citenamefont{Bialynicki-Birula}(2006)}]{bialynicki06}
\bibinfo{author}{\bibfnamefont{I.}~\bibnamefont{Bialynicki-Birula}},
  \bibinfo{journal}{Phys. Rev. A} \textbf{\bibinfo{volume}{74}},
  \bibinfo{eid}{052101} (\bibinfo{year}{2006}).

\bibitem[{\citenamefont{de~Vicente and S\'anchez-Ruiz}(2008)}]{vicente08}
\bibinfo{author}{\bibfnamefont{J.~I.} \bibnamefont{de~Vicente}}
  \bibnamefont{and}
  \bibinfo{author}{\bibfnamefont{J.}~\bibnamefont{S\'anchez-Ruiz}},
  \bibinfo{journal}{Phys. Rev. A} \textbf{\bibinfo{volume}{77}},
  \bibinfo{pages}{042110} (\bibinfo{year}{2008}).

\bibitem[{\citenamefont{Zozor et~al.}(2011)\citenamefont{Zozor, Portesi,
  Sanchez-Moreno, and Dehesa}}]{zozor11}
\bibinfo{author}{\bibfnamefont{S.}~\bibnamefont{Zozor}},
  \bibinfo{author}{\bibfnamefont{M.}~\bibnamefont{Portesi}},
  \bibinfo{author}{\bibfnamefont{P.}~\bibnamefont{S\'anchez-Moreno}},
  \bibnamefont{and} \bibinfo{author}{\bibfnamefont{J.~S.}
  \bibnamefont{Dehesa}}, \bibinfo{journal}{Phys. Rev. A}
  \textbf{\bibinfo{volume}{83}}, \bibinfo{pages}{052107}
  (\bibinfo{year}{2011}).

\bibitem[{\citenamefont{Arthurs and Jr}()}]{Arthurs-Kelly}
\bibinfo{author}{\bibfnamefont{E.}~\bibnamefont{Arthurs}} \bibnamefont{and}
  \bibinfo{author}{\bibfnamefont{J.~L.} \bibnamefont{Kelly}},
\bibinfo{journal}{Bell Syst. Tech.} \textbf{\bibinfo{volume}{44}},
  \bibinfo{pages}{725} (\bibinfo{year}{1965}).

\bibitem[{\citenamefont{Raymer}(1994)}]{Raymer1994}
\bibinfo{author}{\bibfnamefont{M.}~\bibnamefont{Raymer}},
  \bibinfo{journal}{Am. J. Phys.} \textbf{\bibinfo{volume}{62}},
  \bibinfo{pages}{986} (\bibinfo{year}{1994}).
  
{  
   \bibitem[{\citenamefont{Ozawa}(2003)}]{ozawa}
\bibinfo{author}{\bibfnamefont{M.}~\bibnamefont{Ozawa}},
  \bibinfo{journal}{Phys. Rev. A} \textbf{\bibinfo{volume}{67}},
  \bibinfo{pages}{042105} (\bibinfo{year}{2003}).
  
  
  \bibitem[{\citenamefont{Erhart et~al.}(2012)
  \citenamefont{Erhart, Sponar, Sulyok, Badurek, Ozawa, Hasegawa}}]{erhart}
\bibinfo{author}{\bibfnamefont{J.} \bibnamefont{Erhart}},
  \bibinfo{author}{\bibfnamefont{S.}~\bibnamefont{Sponar}},
  \bibinfo{author}{\bibfnamefont{G.} \bibnamefont{Sulyok}},
  \bibinfo{author}{\bibfnamefont{G.}~\bibnamefont{Badurek}}, 
  \bibinfo{author}{\bibfnamefont{M.}~\bibnamefont{Ozawa}},
  \bibnamefont{and}
  \bibinfo{author}{\bibfnamefont{Y.} \bibnamefont{Hasegawa}},
  \bibinfo{journal}{Nature Phys.} \textbf{\bibinfo{volume}{8}},
  \bibinfo{pages}{185} (\bibinfo{year}{2012}).
  
  \bibitem[{\citenamefont{Watanabe et~al.}(2011)\citenamefont{Watanabe, Sagawa, 
and Ueda}}]{watanabe}
\bibinfo{author}{\bibfnamefont{Y.}~\bibnamefont{Watanabe}},
  \bibinfo{author}{\bibfnamefont{T.}~\bibnamefont{Sagawa}},
   \bibnamefont{and}
  \bibinfo{author}{\bibfnamefont{M.}~\bibnamefont{Ueda}}, 
  \bibinfo{journal}{Phys. Rev. A}
  \textbf{\bibinfo{volume}{84}}, \bibinfo{pages}{042121}
  (\bibinfo{year}{2011}).

  }
  

\bibitem[{\citenamefont{Rudnicki et~al.}(2011)\citenamefont{Rudnicki, Walborn,
  and Toscano}}]{Rudnicki2011}
\bibinfo{author}{\bibfnamefont{\L.}~\bibnamefont{Rudnicki}},
  \bibinfo{author}{\bibfnamefont{S.~P.} \bibnamefont{Walborn}},
  \bibnamefont{and} \bibinfo{author}{\bibfnamefont{F.}~\bibnamefont{Toscano}},
  \bibinfo{journal}{Europhys. Lett.} \textbf{\bibinfo{volume}{97}},
  \bibinfo{pages}{38003} \bibinfo{year}{(2012). }

\bibitem[{\citenamefont{Peres}(1995)}]{peres95}
\bibinfo{author}{\bibfnamefont{A.}~\bibnamefont{Peres}},
  \emph{\bibinfo{title}{Quantum Theory: Concepts and Methods}}
  (\bibinfo{publisher}{Kluwer}, \bibinfo{address}{Dordrecht},
  \bibinfo{year}{1995}).

\bibitem[{\citenamefont{Ballentine}(1998)}]{ballentine98}
\bibinfo{author}{\bibfnamefont{L.}~\bibnamefont{Ballentine}},
  \emph{\bibinfo{title}{Quantum Mechanics: A Modern Development}}
  (\bibinfo{publisher}{World Scientific}, \bibinfo{address}{Singapore},
  \bibinfo{year}{1998}).

\bibitem[{\citenamefont{Peres}(1992)}]{peres92}
\bibinfo{author}{\bibfnamefont{A.}~\bibnamefont{Peres}},
  \bibinfo{journal}{Found. Phys.} \textbf{\bibinfo{volume}{22}},
  \bibinfo{pages}{819} (\bibinfo{year}{1992}).

\bibitem[{\citenamefont{Kofler and Brukner}(2007)}]{kofler07}
\bibinfo{author}{\bibfnamefont{J.}~\bibnamefont{Kofler}} \bibnamefont{and}
  \bibinfo{author}{\bibfnamefont{\v{C}.} \bibnamefont{Brukner}},
  \bibinfo{journal}{Phys. Rev. Lett.} \textbf{\bibinfo{volume}{99}},
  \bibinfo{pages}{180403} (\bibinfo{year}{2007}).

\bibitem[{\citenamefont{Zozor and Vignat}(2007)}]{zozor07}
\bibinfo{author}{\bibfnamefont{S.}~\bibnamefont{Zozor}} \bibnamefont{and}
  \bibinfo{author}{\bibfnamefont{C.}~\bibnamefont{Vignat}},
  \bibinfo{journal}{Physica A} \textbf{\bibinfo{volume}{375}},
  \bibinfo{pages}{499} (\bibinfo{year}{2007}).

\bibitem[{\citenamefont{R\'enyi}(1961)}]{renyi61}
\bibinfo{author}{\bibfnamefont{A.}~\bibnamefont{R\'enyi}}
  (\bibinfo{publisher}{Univ. Calif. Press}, \bibinfo{year}{1961}),
  vol.~\bibinfo{volume}{1} of \emph{\bibinfo{series}{Proceedings of the Fourth
  Berkeley Symposium on Mathematical Statistics and Probability}}, pp.
  \bibinfo{pages}{547--561}.

\bibitem[{\citenamefont{Rudin}(1991)}]{Rudin1991}
\bibinfo{author}{\bibfnamefont{W.}~\bibnamefont{Rudin}},
  \emph{\bibinfo{title}{Functional Analysis (2nd ed.)}}
  (\bibinfo{publisher}{McGraw-Hill}, \bibinfo{year}{1991}).

\bibitem[{\citenamefont{Rudnicki}()}]{rudnicki10}
\bibinfo{author}{\bibfnamefont{{\L}.}~\bibnamefont{Rudnicki}},
  \emph{\bibinfo{title}{Uncertainty related to position and momentum
  localization of a quantum state}}, \eprint{arXiv:1010.3269v1} (2010).

\bibitem[{\citenamefont{Abramowitz and Stegun}(1964)}]{abramowitz}
\bibinfo{author}{\bibfnamefont{M.}~\bibnamefont{Abramowitz}} \bibnamefont{and}
  \bibinfo{author}{\bibfnamefont{I.}~\bibnamefont{Stegun}},
  \emph{\bibinfo{title}{Handbook of Mathematical Functions}}
  (\bibinfo{publisher}{Dover}, \bibinfo{address}{New York},
  \bibinfo{year}{1964}).

\bibitem[{\citenamefont{Chafai}(2002)}]{chafai}
\bibinfo{author}{\bibfnamefont{D.}~\bibnamefont{Chafa{\"i}}},
  \bibinfo{journal}{S{\'e}minaire de probabiliti{\'e}s, Strasbourg}
  \textbf{\bibinfo{volume}{36}}, \bibinfo{pages}{194} (\bibinfo{year}{2002}).

\bibitem[{\citenamefont{Riesz}(1927)}]{riesz}
\bibinfo{author}{\bibfnamefont{M.}~\bibnamefont{Riesz}}, \bibinfo{journal}{Acta
  Math.} \textbf{\bibinfo{volume}{49}}, \bibinfo{pages}{465}
  (\bibinfo{year}{1927}).

\bibitem[{\citenamefont{Schurmann}(2008)}]{schur}
\bibinfo{author}{\bibfnamefont{T.}~\bibnamefont{Sch{\"u}rmann}},
  \bibinfo{journal}{Act. Phys. Pol. B} \textbf{\bibinfo{volume}{39}},
  \bibinfo{pages}{587} (\bibinfo{year}{2008}).

\bibitem[{\citenamefont{Bialynicki-Birula and Rudnicki}(2011)}]{bialynicki11}
\bibinfo{author}{\bibfnamefont{I.}~\bibnamefont{Bialynicki-Birula}}
  \bibnamefont{and}
  \bibinfo{author}{\bibfnamefont{{\L}.}~\bibnamefont{Rudnicki}}, \bibinfo{title}{Entropic Uncertainty Relations in Quantum
Physics} in
  \emph{\bibinfo{booktitle}{Statistical Complexity}},
  edited by \bibinfo{editor}{\bibfnamefont{K.~D.} \bibnamefont{Sen}}
  (\bibinfo{publisher}{Springer}, \bibinfo{year}{2011}), p.
  \bibinfo{pages}{1-34}; \eprint{arXiv:1001.4668v1} (2010).

\bibitem[{\citenamefont{Cover and Thomas}(2006)}]{cover}
\bibinfo{author}{\bibnamefont{Cover}} \bibnamefont{and}
  \bibinfo{author}{\bibnamefont{Thomas}}, \emph{\bibinfo{title}{Elements of
  Information Theory}} (\bibinfo{publisher}{John Wiley and Sons},
  \bibinfo{year}{2006}).

\end{thebibliography}

\end{document}